\documentclass[onecolumn,showpacs,preprintnumbers,amsmath,amssymb]{revtex4}

\usepackage{graphicx}% Include figure files
\usepackage{dcolumn}% Align table columns on decimal point
\usepackage{bm}% bold math

\begin{document}

\title{A weighted evolving network model more approach to reality}
\author{Chuan-Ji Fu,Qing Ou,Wen Chen,Bing-Hong Wang}
\email{bhwang@ustc.edu.cn,Fax:+86-551-3603574}
\author{Ying-Di Jin}
\author{Yong-Wei Niu}
\author{Tao Zhou}
\email{zhutou@ustc.edu} \affiliation{
Nonlinear Science Center and Department of Modern Physics,\\
University of Science and Technology of China,\\
Hefei Anhui, 230026, PR China }

\date{\today}

\begin{abstract}

In search of many social and economical systems, it is
found that node strength distribution as well as degree distribution
demonstrate the behavior of power-law with droop-head and heavy-tail.
We present a new model for the growth of weighted networks considering
the connection of nodes with low strengths. Numerical simulations
indicate that this network model yields three power-law distributions
of the node degrees, node strengths and connection weights.
Particularly, the droop-head and heavy-tail effects can be reflected
in the first two ones by this new model.

\end{abstract}


\maketitle

\section{Introduction}

Complex networks have been studied across many fields of science
and society, stimulated by the fact that many systems in nature
can be described by complex networks with nodes representing
individuals or organizations and edges mimicking the interactions
among them\cite{Reviews}. In the past study, there have been a
variety of important properties from the real-life networks as
well as several artificial networks. Notably, it is found that
many real-life networks are scale-free networks, which means that
the degree distributions of these networks follow a power law form
$P(k)\sim k^{-\gamma}$ for large $k$, where $P(k)$ is the
probability that a node in the network is connected to $k$ other
nodes and $\gamma$ is a positive real number determined by the
given network. Since power laws are free of characteristic scale,
such networks are called "scale-free network". Examples of such
networks are numerous: these include the Internet, the World Wide
Web, social networks of acquaintance or other relations between
individuals, metabolic networks, integer networks, food webs,
etc.\cite{Networks}.

From 1999, because of the ubiquity of scale-free networks, much
attention has been focused on how to model scale-free networks.
One of the most well-known model is introduced by Barab\'{a}si and
Albert\cite{BA 99}, which is very similar to Price's
model\cite{Price}. The BA model suggests that two main ingredients
of self-organization of a network in a scale-free structure are
growth and preferential attachment. These point to the facts that
most networks continuously grow by the addition of new vertices,
and new vertices are preferentially attached to existing vertices
with large number of neighbors. However, not all the edges should
be viewed as the same because many real-life networks display
different interaction weights between nodes such as various
scientific collaboration network and ecosystems\cite{Weighted}.
This weighted networks\cite{Barrat1} are usually described
by a matrix $ w_{ij}$ specifying the weight on the edge connecting
the vertexes $i$ and $j$, with $i,j=1, \ldots, N$, where $N$ is
the size of the network. For simplicity, we only consider
undirected network models so that the weight be merely viewed
symmetrically $w_{ij}=w_{ji}$. Then the strength of the node $i$
can be defined as\cite{Barrat2,Yook}:
\begin{equation}
s_{i}=\sum_{j\in V(i)}w_{ij}
\end{equation}
where the sum runs over the set $V(i)$ of neighbors of $i$. What
should be pointed out is that in most cases the empirical
distribution of $s$ has the behavior of power-law with droop-head
and heavy-tail\cite{Yook}, analogous to the power-law decay of the
degree distribution in weighted networks, while in unweighted
networks the empirical distribution of degree also has the
droop-head and heavy-tail\cite{BA 99}.

Recently, Barrat {\it et al.} proposed a weighted evolving
network\cite{Barrat1}, which recovers an effective preferential
attachment. Their model indeed demonstrates the property of
scale-free, but it loses the properties of droop-head and
heavy-tail. In the present paper, we propose a new model for the
growth of weighted networks considering the connection of nodes
with low strengths. Numerical simulations indicate that this
network model yields three power-law distributions of the node
degrees, node strengths and connection weights. Particularly, the
droop-head and heavy-tail effects can be reflected in the first
two ones by this new model.

The outline of this paper is as follows. In Sec. $\textrm{2}$, we define
the model. In Sec. 3, we report the experimental results.
Finally, in Sec. 4, we give our conclusion and a short
discussion.

\section{The new model}

The proposed model is defined as the following scheme:

First, start from an initial seed of $N_{0}$ nodes connected by
links with assigned weight $w_{0}$. Next, at each time step, a new
node $n$ is added with $m$ edges that are randomly attached to a
previously existing node $i$ according to the probability
distribution
\begin{equation}
\Pi_{n\rightarrow i}=\frac{s_{i}}{\sum_{j}s_{i}}
\end{equation}
This rule relaxes the usual degree preferential attachment,
focusing on a strength driven attachment in which new nodes
connect more likely to nodes handling larger weights and which are
more central in terms of the strength of interactions. The weight
of each new edge is fixed to a value $ w_{0}$. Moreover, the
presence of the new edge $(n,i)$ will introduce variations of the
existing weights across the network. In particular, we consider
the local rearrangements of weights between $i$ and its neighbors
$ j\in V(i)$ according to the simple rule
\begin{equation}
w_{ij}\longrightarrow w_{ij}+\triangle w_{ij}
\end{equation}
where
\begin{equation}
\triangle
w_{ij}=\delta\frac{w_{ij}}{s_{i}}
\end{equation}
and
\begin{equation}
s_{i}\longrightarrow s_{i}+\delta+w_{0}
\end{equation}
which induces a total increase of traffic $\delta$\cite{Barrat1}.
In this paper, we only focus on the simplest model with $ \delta$
a constant(See Fig.1 a).

After given time steps of adding new nodes, the strengths of
different nodes diverse greatly to some extend, which can be
viewed as the diversities of competitive power of the nodes. It
is natural to introduce the rule that the nodes with low strengths
prefer to cooperate with one another in order to increase their
competitive power. Therefore, in the following evolving
process, our model not only add new nodes into the network, but
also permit old nodes to evolve with the growth of the network.
Define a threshold of strength $s_{c}$ to tell the nodes with low
strengths -consisting of the set $G$ of "struggling nodes"- apart
from the network. This means that when $s_{i}<s_{c}$ viz. $ i\in
G$, the node $ i$ struggles to connect with another node $j\in G$
with a given probability $p$ unless there is an existing edge
between $i$ and $j$. (See Fig.1 b) After the connections among old
nodes have been done, the growth process is iterated by
introducing a new node with the corresponding judgment of and
connection among old nodes.

\begin{figure}[!h]
\label{fig1}
\scalebox{0.7}[0.65]{\includegraphics{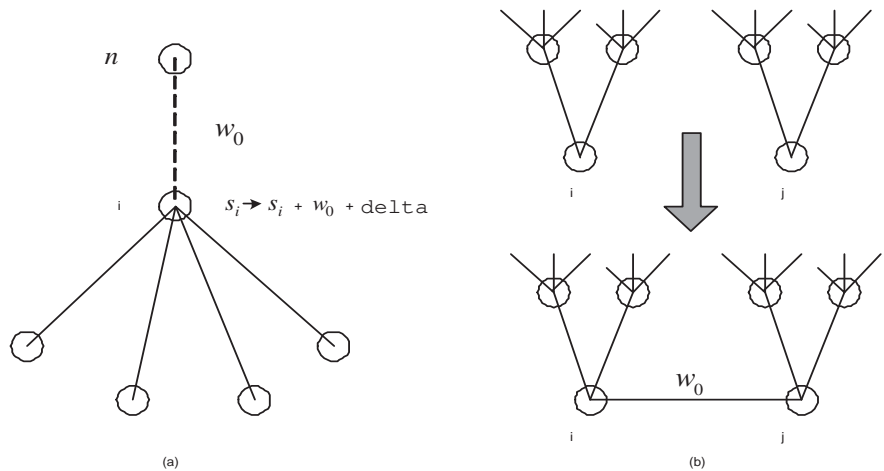}} \\
\caption{(a) Illustration of the mechanism of introducing a new node.
This introduction obeys the strength preferential attachment, and the
strengths and the weights should be rearranged as above. (b)
Illustration of the mechanism of the connection between old nodes.
The nodes with low strengths prefer to connect with each other by
an edge with the weight of $w_{0}$.}
\end{figure}

It is agreeable with the reality that the new model contains the
mechanics of connection among the nodes with low strength. In the
scientific collaboration networks and actor collaboration networks
etc., there exists a common phenomena that the people, whose
competitive abilities are limited, prefer to collaborate with one
another or consist of a team so that they can survive in the
environment full of competitions. Take the combination of
corporations for another example in reality. Small corporations
sometimes have to corporate with one another to increase their
competitions too. Therefore, the mechanism of connection among old
nodes has its root in reality.

\section{Experimental Results}

We performed numerical simulations. For the sake of simplicity, we
set $w_{0}=1$.

In Fig. 2, we show that the node strength distribution $P(s)$
obeys a power-law distribution with obvious droop-head and
heavy-tail which coincides with the statistical results of many
real-life networks\cite{Barrat2}.

\begin{figure}[!h]
\label{fig2}
\scalebox{0.7}[0.65]{\includegraphics{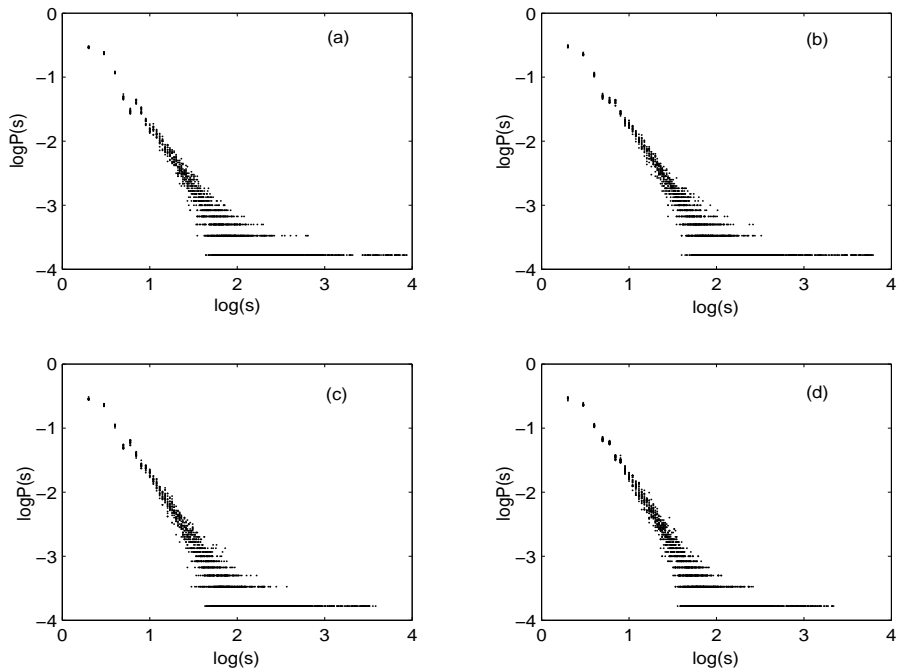}} \\
\caption{The node strength distribution with different
parameters $\delta$ and $p$:
(a) $\delta=2.5$ and $p=0.25$
(b) $\delta=2$ and $p=0.25$
(c) $\delta=1.5$ and $p=0.29$
(d) $\delta=1$ and $p=0.29$.
The data are accumulated over 20 networks of size $N=6000$.
All the illustrations display the power law property with
the obvious droop-head and heavy-tail.}
\end{figure}

In Fig.3, we show that the degree distribution $P(k)$ also obeys a
power-law distribution with droop-head or heavy-tail. This is
again consistent with the statistical results of real-life
networks.

\begin{figure}[!htbp]
\label{fig3}
\scalebox{0.7}[0.65]{\includegraphics{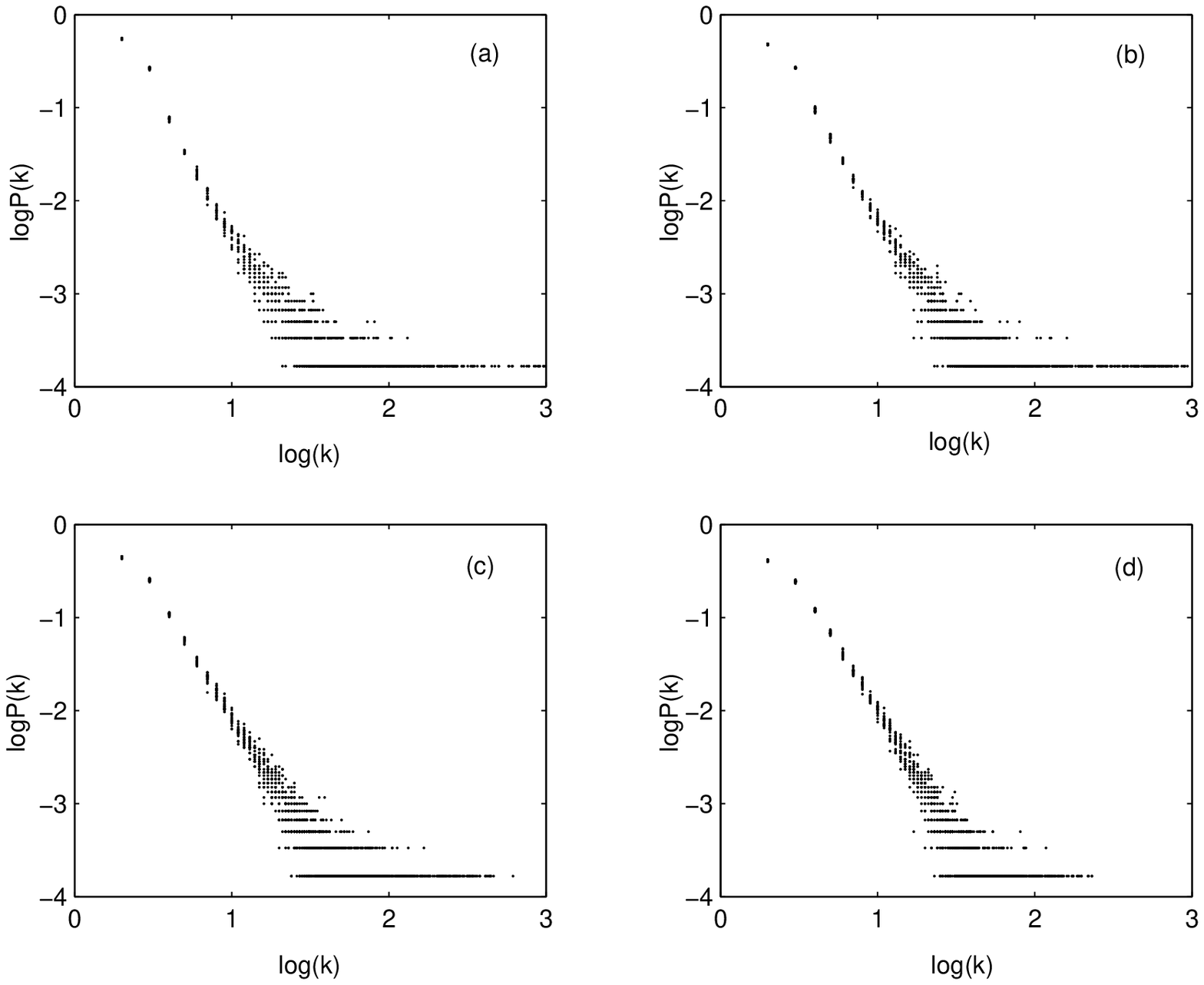}} \\
\caption{The degree distribution with different parameters $\delta$ and $p$:
(a) $\delta=3$ and $p=0.17$
(b) $\delta=1.5$ and $p=0.17$
(c) $\delta=0.5$ and $p=0.13$
(d) $\delta=0$ and $p=0.1$.
The data are accumulated over 20 networks of size $N=6000$.
All the illustrations display the power law property with
the obvious droop-head and heavy-tail. }
\end{figure}

In Fig. 4, we show the edge weight distribution $ P(w)$ obeys
power-law without obvious droop-head and heavy tail that is
similar to many real-life data\cite{Chen}.

\begin{figure}[!htbp]
\label{fig4}
\scalebox{0.7}[0.65]{\includegraphics{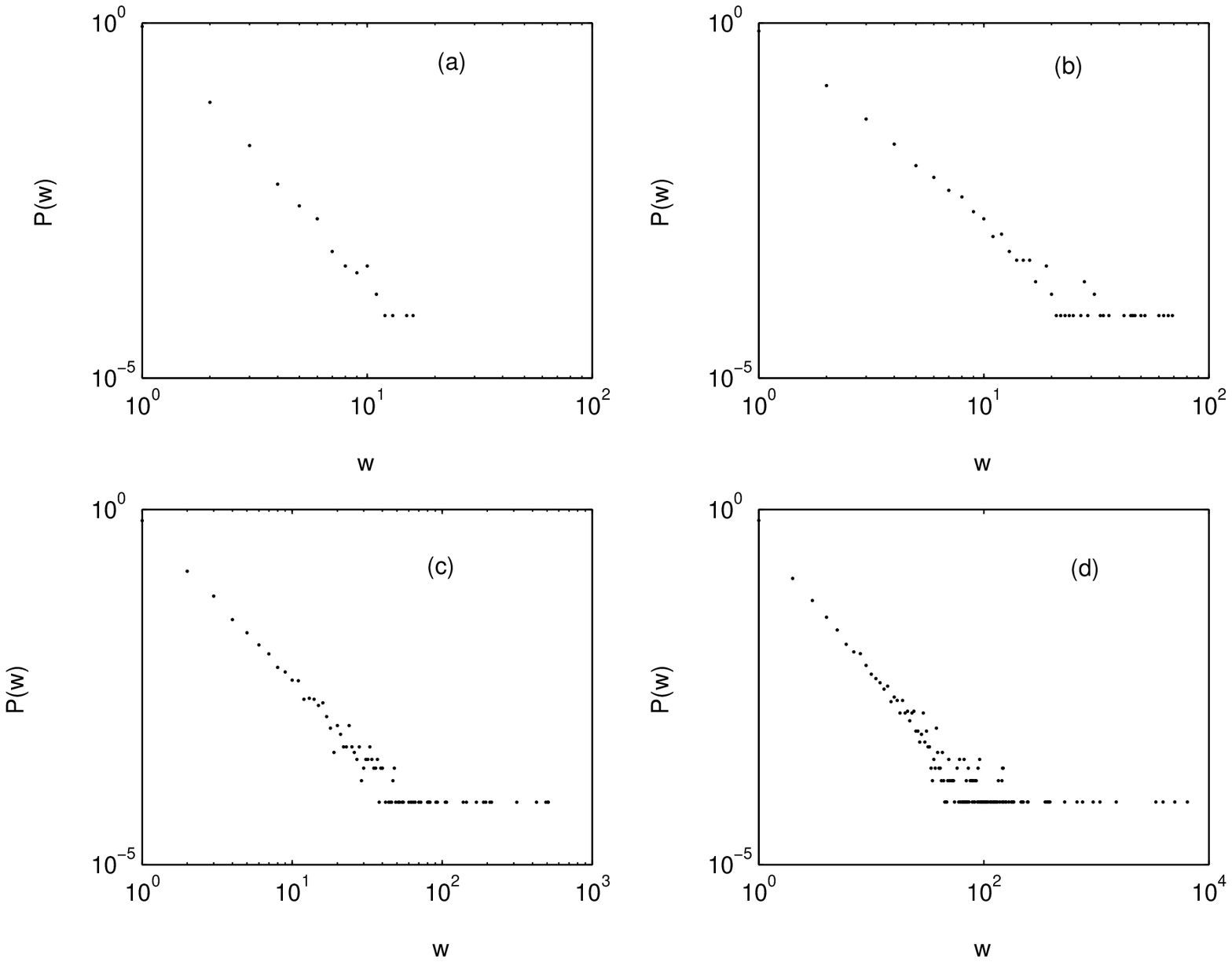}} \\
\caption{ The edge weight distribution with different
parameters $\delta$ and $p$:
(a) $\delta=0.5$ and $p=0.2$
(b) $\delta=1$ and $p=0.2$
(c) $\delta=2$ and $p=0.2$
(d) $\delta=5$ and $p=0.2$.
The data are accumulated over 1 networks of size $N=6000$.
Compared with the strength and degree distributions, weight
distribution obeys relatively more strict power law.}
\end{figure}

In Fig. 5, we show the time evolution of the strengths of three
initial nodes and the weights of two initial edges.

As can be seen, both the $s_{i}(t)$ and the $w_{j}(t)$ increase
linearly with $t$, when $t$ is large enough.

\begin{figure}[!htbp]
\label{fig5}
\scalebox{0.7}[0.3]{\includegraphics{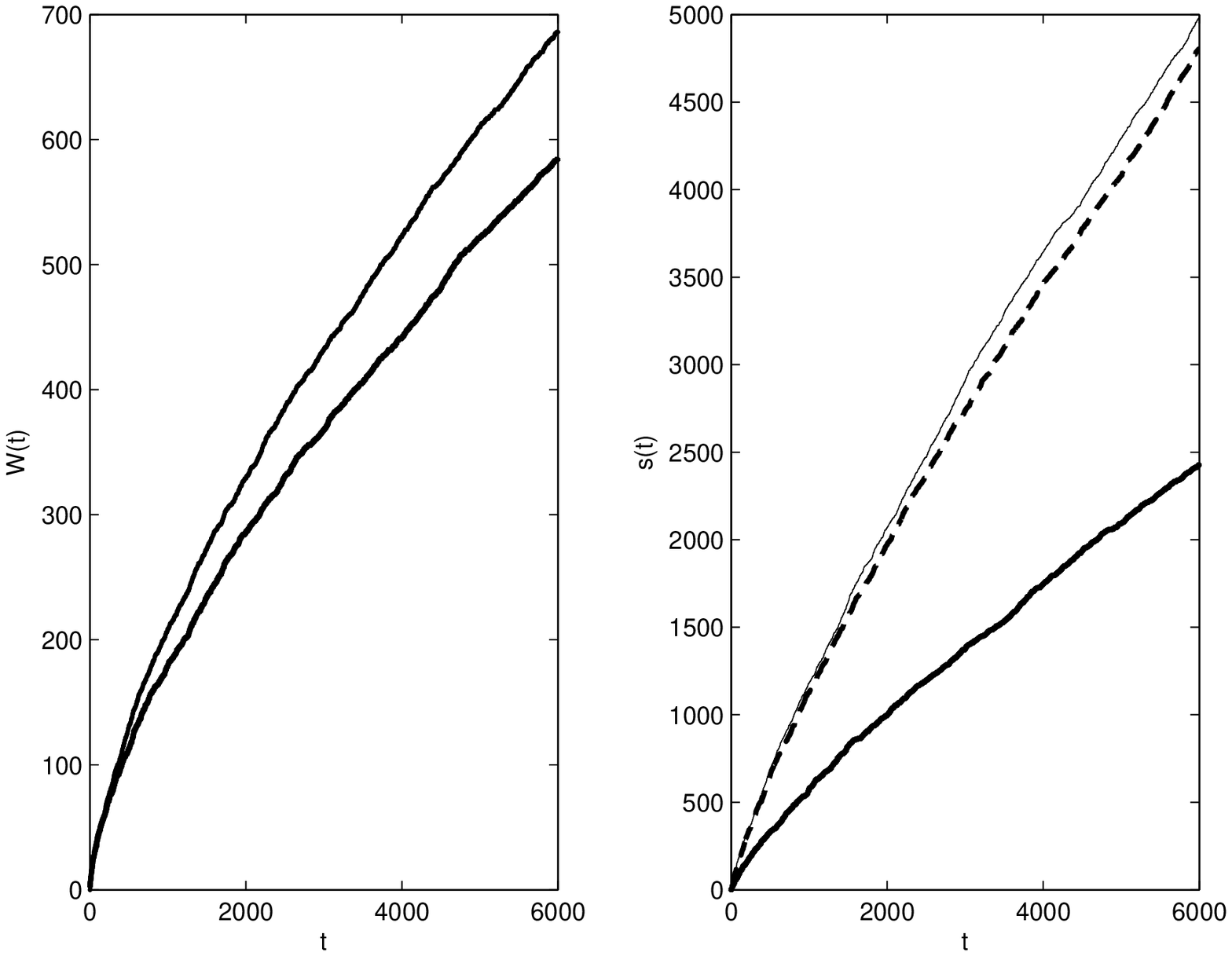}} \\
\caption{The left of Fig.5. illustrates the time evolution of the weights of
two edges with the parameters $\delta=2$ and $p=0.2$.
The right illustrates the evolution of three strengths with the same parameters.
As it can be seen, when $t$ is large enough, both the $s_{i}(t)$
and the $w_{j}(t)$ increase linearly with $t$.}
\end{figure}

\section{conclusion and discussion}
After the weighted evolving model of Barrat\cite{Barrat1}, we
introduce a new one more approach to reality. We
investigated the cause of presence of droop-head and heavy-tail,
which gives a clearly physical picture of how a network evolves. A
more meaningful thought of the model is that it reveals the
mechanism of competition in some social networks. This may shed
some new light on the development of modelling the real social
networks.

There exist a series of modifications that are allowed. First,
$\delta$ being constant is not reasonable. Since not every time step
the new node introduces the same amount of $\gamma$ into the network.
Second, the threshold of strength $s_{c}$ should be replaced by a
function of $s$.
Third, it is worthy to study the property of the spectral density
of this model. Forth, the degree-degree and strength-strength
auto-correlations and degree-strength cross-correlation are also worth
studying.

\begin{acknowledgements}
This work has been partially supported by the State Key
Development Programme of Basic Research (973 Project) of China,
the National Natural Science Foundation of China under Grant
No.70271070 and the Specialized Research Fund for the Doctoral
Program of Higher Education (SRFDP No.20020358009)
\end{acknowledgements}


\begin{thebibliography}{Reviews}

\bibitem{Reviews} R. Albert and A. -L. Barab\'{a}si, Rev. Mod. Phys. {\bf 74}, 47 (2002); S. N. Dorogovtsev and J. F. F. Mendes, Adv. Phys. {\bf 51}, 1079 (2002); M. E. J. Newman, Preprint cond-mat/0303516 (2003).

\bibitem{Networks} A. Vazquez, R. Pastor-Satorras and A. Vespignani, Preprint cond-mat/0303516 (2003); R. Albert, H. Jeong and A. -L. Barab\'{a}si, Nature {\bf 401}, 130 (1999); H. Jeong, B. Tombor, R. Albert, Z. N. Oltvai and A. -L. Barab\'{a}si, Nature {\bf 407}, 651 (2000); J. M. Montya and R. V. Sol\'{e}, J. Theor. Biol. {\bf 214}, 405 (2002); T. Zhou, B. -H. Wang, P. -Q. Jiang, Y. -B. Xie and S. -L. Bu, Preprint cond-mat/0405258 (2004).

\bibitem{BA 99} A. -L. Barab\'{a}si and R. Albert, Science {\bf 286}, 509 (1999); A. -L. Barab\'{a}si, R. Albert and H. Jeong, Physica {\bf A272}, 173 (1999).

\bibitem{Price} D. J. de S. Price, Science {\bf 149}, 510 (1965);
J. Amer. Soc. Inform. Sci. {\bf 27}, 292 (1976).

\bibitem{Weighted} S.L. Pimm, {\it Food Webs} (University of Chicago Press,Chicago,2002), 2nd
ed; A.E. Krause, K.A. Frank, D.M. Mason, R.E. Ulanowicz, and W.W.
Taylor, Nature {\bf426}, 282 (2003).

\bibitem{Barrat1} A. Barrat, M. Barth\'elemy, and A. Vespignani, Phys. Rev. Lett. {\bf92}, 228701 (2004).

\bibitem{Barrat2} A. Barrat, M. Barth\'elemy, R. Pastor-Satorras, and A. Vespignani, Proc. Natl. Acad. Sci. U.S.A. {\bf101}, 3747 (2004).

\bibitem{Yook} S.H. Yook, H. Jeong, A.-L. Barab\'asi, and Y. Tu, Phys. Rev. Lett. {\bf86}, 5835 (2001).

\bibitem{Chen} C. Li and G. Chen, Preprint cond-mat/0311333 (2003).

\end{thebibliography}
\end{document}